# Tests Of The Charged Particle Stepper With Muons

C. Milsténe[a,b], G. Fisk[a], A. Para[a]

[a]Fermi National Laboratory, Batavia Illinois
[b]Northern Illinois University, DeKalb Illinois

Version 2-07-05

# Introduction

In the initial simulation of ALCPG muon tracking, the projection of a muon, or other charged particle, through the detector assumes a helical path without corrections for energy loss. Since the generation of Monte Carlo events using GEANT properly takes into account multiple scattering, energy loss, etc. there are systematic differences in comparisons of the projected track with the generated track, especially for low momentum muons. The helical swimmer used so far in simulation studies is an excellent tool for a tracker in which the material density is low. However, while following charged particles through the calorimeters, the coil and the muon detector the loss of energy in the material must be taken into account. The long lever arm of the Muon Detector stresses the necessity of a program which accounts for the loss of energy of the particle in the material. The stepper answers that need and is discussed in this note. We first present an outline of the algorithm of the stepper that is then tested on the Muon Analysis code. The inclusion into the Muon code and the improvement for track matching in the hadron calorimeter and muon detector are shown.

# Algorithm Outline

The stepper is performing a chosen number of steps through the material located in the way of the particles, e.g. in the calorimeters layers, and the loss by dE/dx as well as the change in direction due to the magnetic field are reflected upon the momentum change, each step, and the position and momentum recalculated anew for the next step.
The motion of a charged particle in a uniform magnetic field $B_z$ with energy loss in the medium can be expressed by a change in direction of the momentum in the ($p_x$, $p_y$) plane due to the magnetic field, followed by a change in magnitude of momenta ($p_x$, $p_y$, $p_z$) due to the loss in energy in the material.
One starts with a particle at the interaction point (IP), at a given position x, y, z ~ 0, 0, 0 with a given momentum $p_x$, $p_y$, $p_z$ and mass. The motion through matter in a magnetic field is given by:

$$p_x(n+1) = p_x(n) + 0.3*q*\frac{P_y(n)}{E(n)}*clight*B_z*\Delta t(n) + \gamma_x(n)$$

$$p_y(n+1) = p_y(n) - 0.3*q*\frac{P_x(n)}{E(n)}*clight*B_z*\Delta t(n) + \gamma_y(n)$$

$$p_z(n+1) = p_y(n) + \gamma_z(n)$$

The 2$^{nd}$ term in $p_x$ and $p_y$ is the usual qv×B term due to the field Bz, and the 3$^{rd}$ term comes from energy loss. Because the field is in the z direction, $p_z$ changes only due to the energy loss in the media. Here $p_x(n)$, $p_y(n)$, $p_z(n)$ are in GeV/c, E(n) in GeV and clight=3E08 m/s, is the velocity of light, $\Delta t(n)$ the time duration between steps is in seconds.

Change in energy

Due to its motion the particle loses energy in the material and therefore slows down accordingly, with a change in the kinetic energy .In step n, for a path length $\Delta s$, assuming that dE/dx = Constante = Ct and given by the hep library

$$\Delta E = (\frac{dE}{dx})*\Delta s = Ct*\Delta s$$

$$\Delta E = \frac{dE}{dP}*\Delta P = \frac{P(n)}{E(n)}*\Delta P$$

Change in momentum due to the passage through matter between media boundaries
From the two previous equations one gets

$$\frac{P(n)}{E(n)}*\Delta P = Ct*\Delta s$$

$$\Delta P = \frac{E(n)}{P(n)}*Ct*\Delta s$$

The results of the calculation is given below, the detail of the calculations is in the appendix.

$$\gamma_x(n) = \Delta p_x = E(n)*(\frac{dE}{dx})*\frac{p_x(n)}{P^2(n)}*\Delta s = E(n)*(\frac{dE}{dx})*\frac{p_x(n)}{p_x^2(n) + p_y^2(n) + p_z^2(n)}*\Delta s$$

$$\gamma_y(n) = \Delta p_y = E(n)*(\frac{dE}{dx})*\frac{p_y(n)}{P^2(n)}*\Delta s = E(n)*(\frac{dE}{dx})*\frac{p_y(n)}{p_x^2(n) + p_y^2(n) + p_z^2(n)}*\Delta s$$

$$\gamma_z(n) = \Delta p_z = E(n)*(\frac{dE}{dx})*\frac{p_z(n)}{P^2(n)}*\Delta s = E(n)*(\frac{dE}{dx})*\frac{p_z(n)}{p_x^2(n) + p_y^2(n) + p_z^2(n)}*\Delta s$$

The momentum decreases due to loss to matter, and therefore the $\gamma_x$, $\gamma_y$, $\gamma_z$ are negative quantities.

One notices that the change in momentum is completely symmetrical between the components $p_x$ and $p_y$. The bigger the component the bigger the loss in that direction, as is expected.
There is also a component $p_z$ to which only the passage through matter contributes, which has the same form.

Path

The position $x(n+1)$, $y(n+1)$, $z(n+1)$ is recalculated after each step as a function of the new values of px, py, pz ,E and the old position $x(n)$, $y(n)$, $z(n)$ as follow:

$$x(n+1) = x(n) + \frac{p_x(n)}{E(n)} * clight * \Delta t(n)$$

$$y(n+1) = y(n) + \frac{p_y(n)}{E(n)} * clight * \Delta t(n)$$

$$z(n+1) = z(n) + \frac{p_z(n)}{E(n)} * clight * \Delta t(n)$$

and $\Delta t(n)$ is the time of flight in seconds of the particle at step n and is calculated below as a function of $p_i(n)$, i = x, y, z, E(n), and x(n),y(n),z(n).
The position at step (n+1) is given in cm, therefore we express clight =3E10 cm/s. The momentum components are given in GeV/c, and the energy in GeV as before

Time of flight (that step)

Is calculated by looking at the radii $r(n+1)$ and $r(n)$ in the (x, y) plane, $r(n)$ being the radius of the layer crossed by the particle in step n and the $V_i(n)$ is given by

$$V_i(n) = \frac{p_i(n)}{E(n)} c; i = x, y, z$$

Between steps n and n+1 one has

$$[r(n+1)]^2 - [r(n)]^2 = ([x(n+1)]^2 + [y(n+1)]^2) - ([x(n)]^2 + [y(n)]^2);$$
$$= [x(n) + v_x(n) * \Delta T(n)]^2 + [y(n) + v_y(n) * \Delta T(n)]^2 - ([x(n)]^2 + [y(n)]^2);$$
$$= 2 * [x(n) * v_x(n) + y(n) * v_y(n)] * \Delta T(n) + ([v_x(n)]^2 + [v_y(n)]^2) * [\Delta T(n)]^2;$$

$dd = StepSize = $ Thickness of the layer/N; N = Number of steps
$$[r(n+1)]^2 = [r(n)]^2 + 2 * dd * \sqrt{([x(n)]^2 + [y(n)]^2} + dd^2$$
$$c1 = [r(n+1)]^2 - [r(n)]^2 = 2 * dd * \sqrt{[x(n)]^2 + [y(n)]^2} + dd^2$$
and $\Delta T(n)$ is given below as the solution of an equation of the 2$^{nd}$ order.

$$\Delta T(n) = \frac{-b + \sqrt{b^2 - 4*a*c}}{2*a}; \quad c = -[2*d*r(n) + d^2]$$
$$a = v_x^2(n) + v_y^2(n); \quad b = 2*[x(n)*v_x(n) + y(n)*v_y(n)]$$

Following is a simple processing flow diagram of the algorithm.

# The Processing Flow

## Stepper Processing Flow

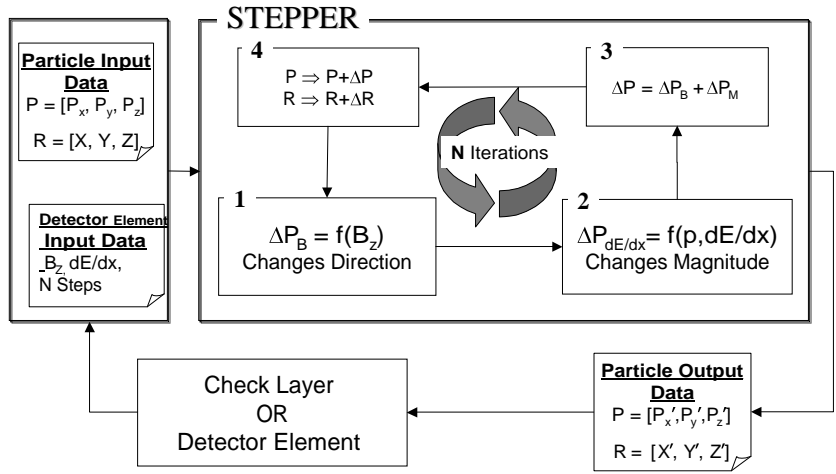

C. Milsténe

**Figure 1- The Stepper – 4 stages iterated N times between Input and Output.**

The flow diagram given above represents the processes involved in the stepper. The following is a list of the inputs, processes, and outputs from the stepper routine.
 Input (a),(b)
a) Particle Input Data: {x,y,z,$p_x$,$p_y$,$p_z$,m,q}  b) Detector Element Input Data: $B_z$,<dE/dx>,N Steps
   Where <dE/dx> is the mean value that layer
  Stepper –{ (1),(2),(3),(4) }× N times
   1) $\Delta P_B$ - changes direction due to the field Bz
   2) $\Delta P_{dE/dx}$ – changes magnitude due to dE/dx
   3) $\Delta P = \Delta P_B + \Delta P_{dE/dx}$
   4) P=P+$\Delta$P; R=R+$\Delta$R
Output: {x', y', z', $p_x$', $p_y$', $p_z$', m, q), will become eventually the new Input Data
For the next layer or the next Detector element.

# The Package Used

The original Muon package used was generated at NIU by R. Markeloff [1] .C. Milstene extended it to include the EMCal barrel. This package contains the java code below which should be compiled in the order they are here.

    SegmentFinder.java
    BarrelCalSegmentFinder.java
    MuonCalSegmentFinder.java
    which contains an inner class:
    TrackExtrapolator.java
    MuonCandidate.java
    MuonList.java
    MuonReco.java

The classes produced should be located in a folder named
hep\lcd\recon\muon\
The above folder has to be included in another folder containing the driver:
LCDMuonDriver.java

# Code Modification to Include the Stepper

The Track Stepper class follows exactly the algorithm presented in the outlines.
The following sub-directory has to be created, and it must contain the stepper class:

Hep\lcd\util\step\

And one needs to add an import statement in both classes
BarrelCalSegmentFinder.java and
MuonCalSegmentFinder.java methods

import hep.lcd.util.step

In the MuonCalSegmentFinder.java, the actual steps through the material replace the track extrapolation; therefore, in MuonCalSegmentFinder.java only about one-third of The original code is left, the following class has disappeared:.

private class TrackExtrapolator for the BarrelCalSegmentFinder instance field, where the object *swmr* was defined one also defines the object *stpr*, as shown below:

protected HelicalSwimmer *swmr* = new HelicalSwimmer( );
protected TrackStepper *stpr* = new TrackStepper( ); //C.M.-

Whenever a call to the helical swimmer was used in BarrelCalSegmentFinder.java or MuonCalSegmentFinder.java, namely.

*swmr*.swim(params, r, zmax);
rint = *swmr*.getIntersect( );
 is replaced by

*stpr*.tkSteps(r, zmax, stepConditions); and
rstep  = *stpr*.getRAfterStep( );

Where rstep in the stepper replaces rint in the swimmer and represents the coordinates of the last point reached by the particle.

In the case of the stepper one requires that each segment of material encountered is stepped with the correct loss of energy by dE/dx in the material and the correct magnetic field. This includes the layers of air in between the calorimeters as well as the 9.2 cm of carbon left in between the EMCAL and the HCAL for the SiD –LC detector. In the calorimeter, when the charged particle reaches the center of the last layer an extra step of half a layer, absent for the swimmer, is required for the stepper.

On the other hand, to go through the tracker up to the entrance of EMCAL the swimmer of Ray Frey ("Javaized" by Tony Johnson) and the stepper give the same results. The last point reached has the exact same coordinates such that the swimmer and the stepper could be interchanged when there is very little/or no material, as in the tracker.

# Improvements in HCAL and MUDET

For particles above ~20 GeV/c the swimmer properly represents the hits, but at lower energy the effect of the energy loss on the particle trajectory is important. The mean energy lost by dE/dx by the particle from the Interaction Point (IP) to the MUDET is ~1.2GeV/c.
Below is shown the overlay of the hits and the extrapolated tracks layer by layer as a function of the layer number for the same 3 GeV/c Muon, with the swimmer and also with the Stepper, which includes dE/dx. One notices that in the last layers at the end of HCAL there is no overlap of hits with the swimmer whereas there is excellent overlap with the stepper.
At higher energy the muon detection efficiency is somewhat higher with the stepper than with the swimmer, but the discrepancies between swimmer and stepper are less important. The higher the energy the closer are the results with both methods, e.g. the agreement is better at 20 GeV/c than at 10 GeV/c. Therefore we have extensively tested the region between 3GeV and 5 GeV .
The agreement is also worse farther away from the interaction point. Therefore at low momenta the results are better with HCAL than with MUDET.
Below is represented the track extrapolation in HCAL using the swimmer versus the hits. One can see that around the middle of the track the two curves are separate from each other. The next plot shows the track extrapolation using the stepper with the loss of energy by dE/dx included versus the hits. Here one sees that both the track extrapolation and the hits are undistinguishable.

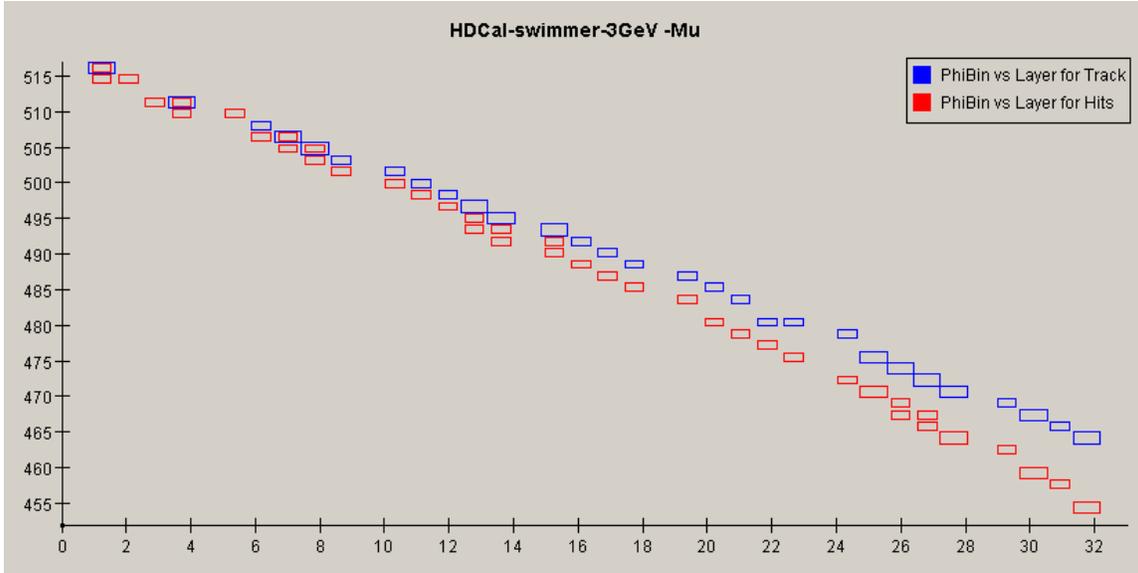

**Figure 2- HCal : Track From the Swimmer vs. Hits- Φ Bin= f (Layer Number)  - Total:1200 Bins, 34 Layers**

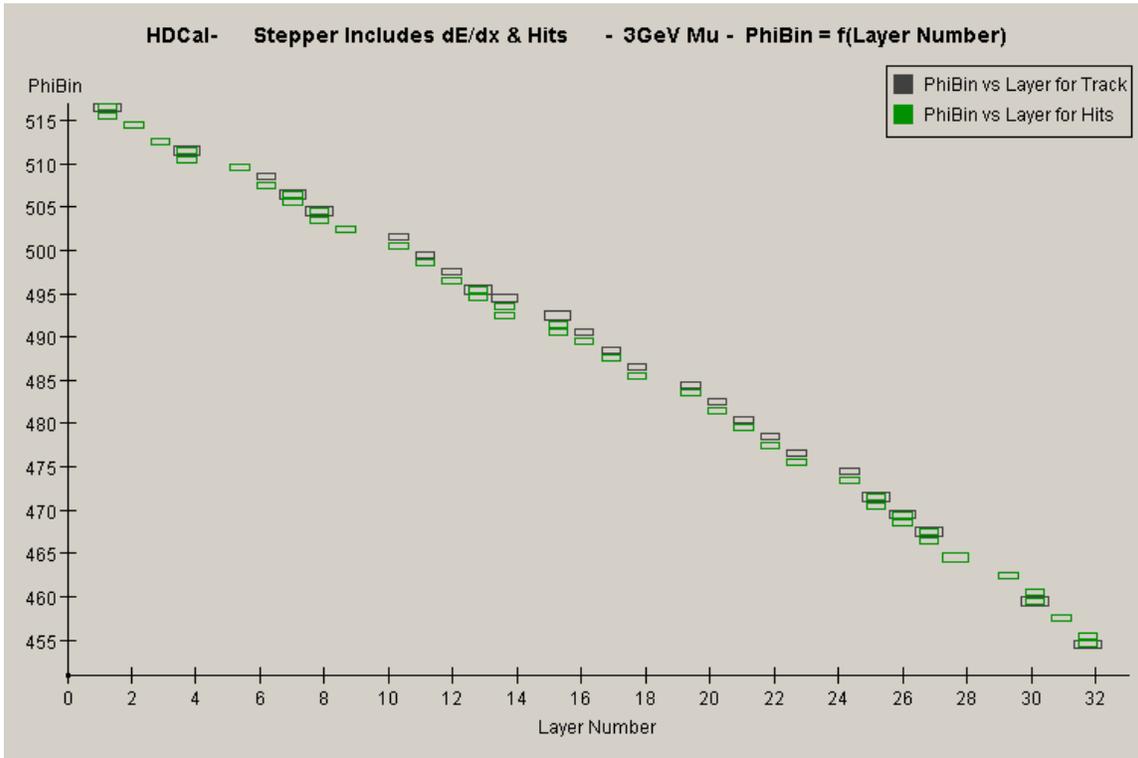

**Figure 3- HCal: Track From The Stepper vs. Hits - Φ Bin = f ( Layer Number)  – Total: 1200 Bins, 34 Layers**

In Figure 4 is represented the overlay of hits and tracks in the MUDET layer by layer as a function of the layer number with the swimmer. Once can see the discrepancy is bigger than in HCAL.

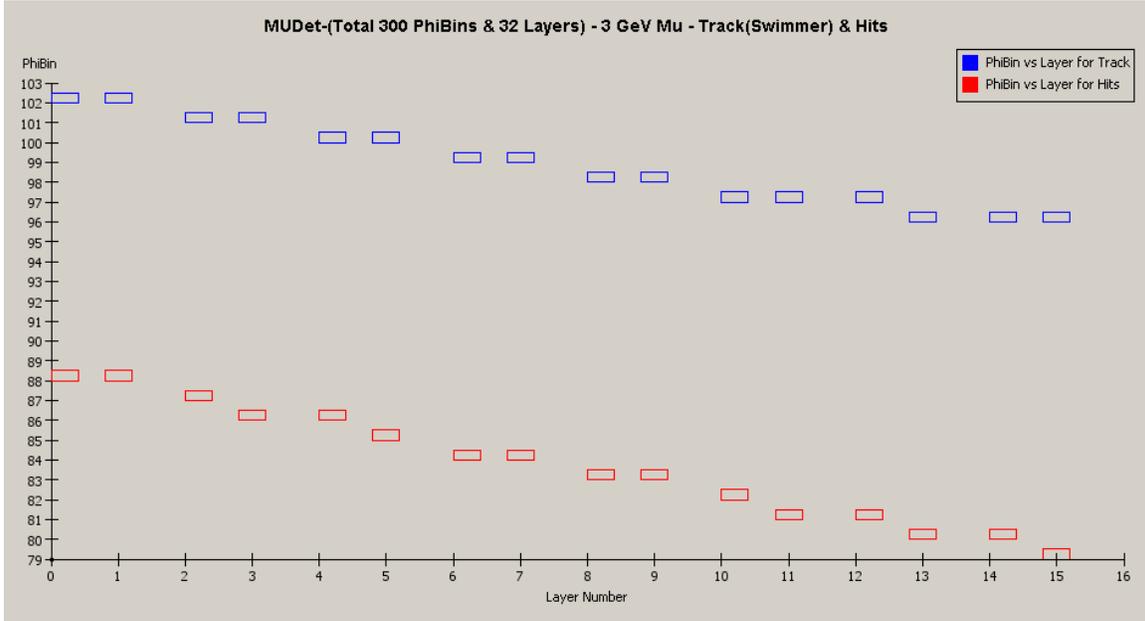

**Figure 4- MuDet Track with the Swimmer &Corresponding Generated Hits.
ΦBin = f (Layer Number) - 300 Bins 32 Layers.**

Then after using the stepper and a magnetic field which drops from 5 Tesla to –0.6 Tesla at the entrance of the coil and up to a radius of 550cm one gets the overlay hits track layer by layer as a function of the layer which, even at 3GeV/c, shows a very good match in MUDET , shown in the next figure, same event as in the previous figure.

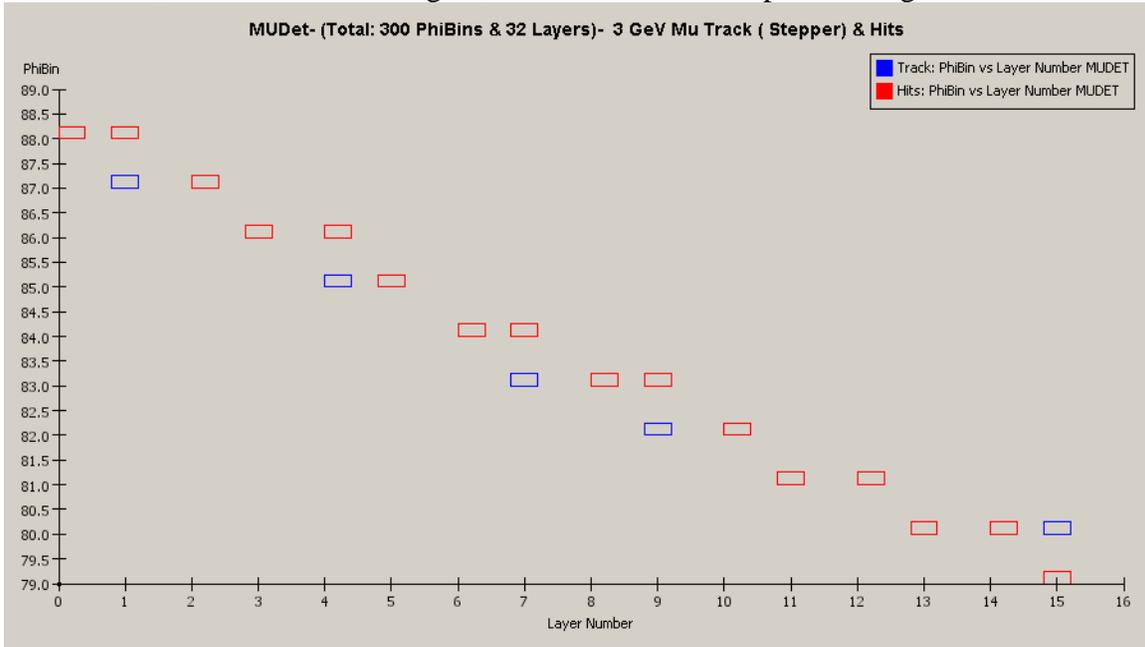

**Figure 5 - Same Mudet Track with the Stepper & Corresponding Hits. Track and Hits fall on top of each other**

In the figure below the Track from the Stepper and the corresponding Generated Hits are shown in MUDET as well as in EMCAL and HCAL

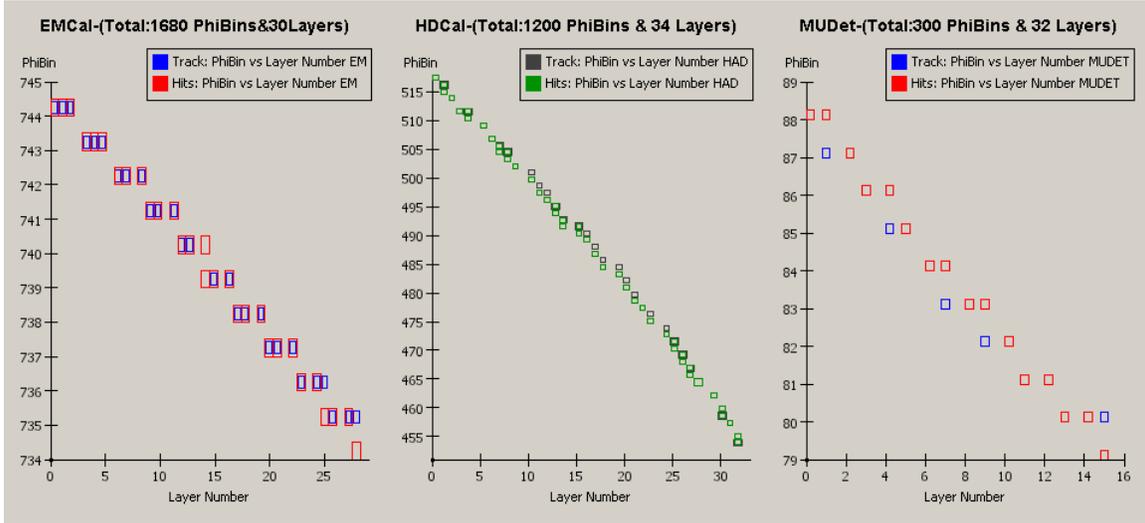

**Figure 6- EMCal, HCal, MuDet - Track with the Stepper Which Includes dE/dx And Generated Hits**

   **Φ Bin Number = f ( Layer Number)**

Below is represented in Figure 6, the distribution of the difference in Phi between tracks and hits for layers 0, 4, 6, 12, 16, 22, 26 and 29 in the Hadron Calorimeter (HCAL). The ΔΦ(Track-Hit) is relatively well centered on the zero. These distributions are done at 4GeV/c.

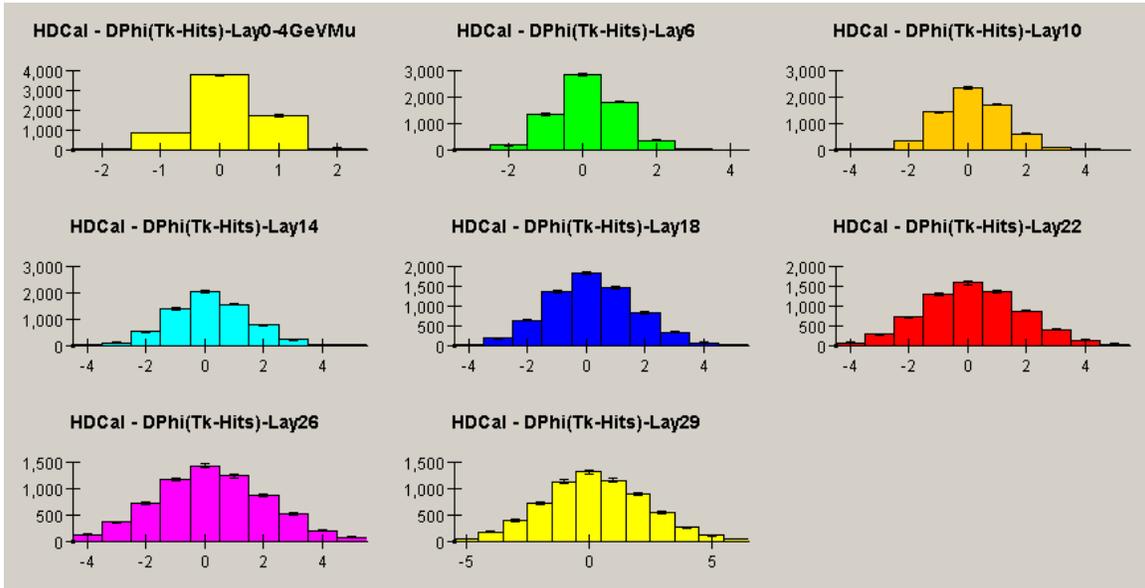

**Figure 7-Hadron Calorimeter: The Distribution |ΦTrack -Φ Hit |, For Layers 0, 4, 6, 12, 16, 22, 26, 29**

In Figure 7, the same distribution is shown in the Muon Detector ( MUDET) for 4GeV Muons . It is centered at zero in the central layers of MUDET but have a very slight shift at the borders of the detector.

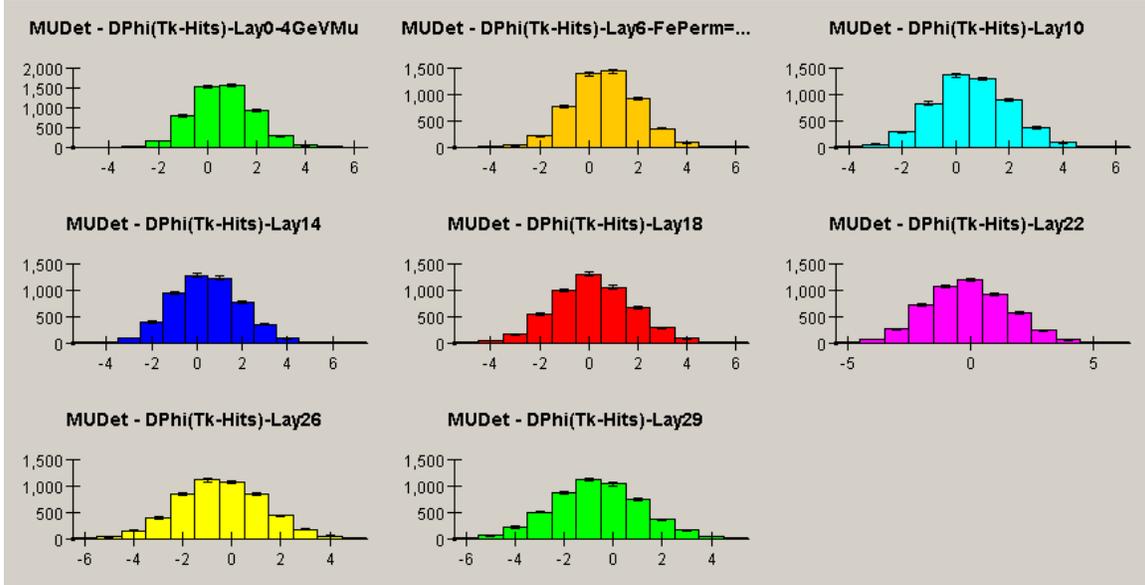

**Figure 8- Muon Detector: The Distribution |ΦTrack - Φ Hit|, For Layers 0, 6, 10, 14, 18, 22, 26, 29**

We are also showing, in Figure 8, the ΔΘ distribution of the 4 GeV Muons in MUDET. There too the distribution is centered at zero.

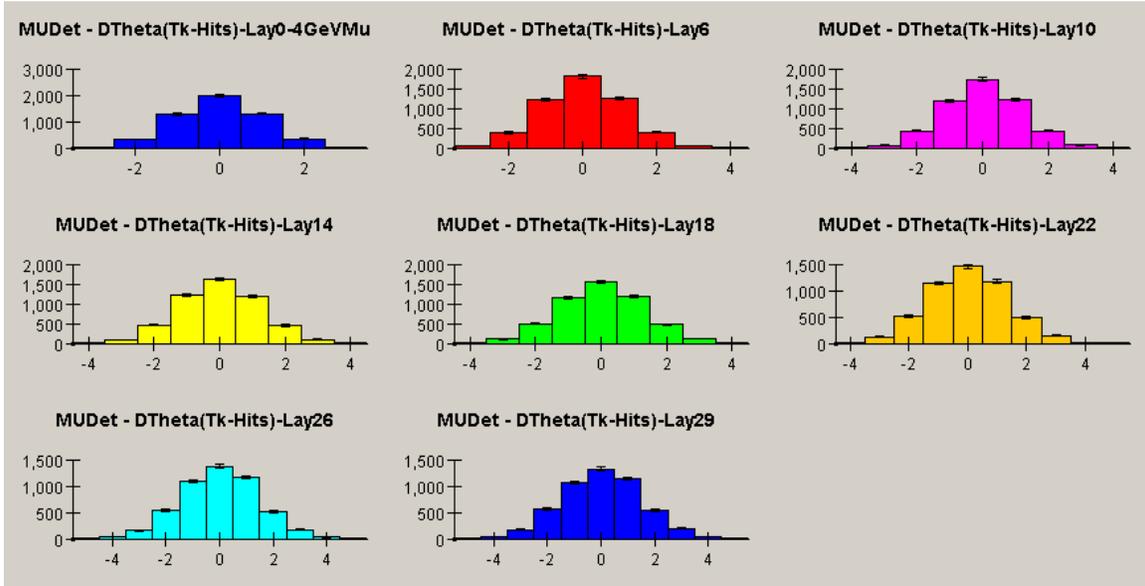

**Figure 9- Muon Detector: The Distribution |The Track - The Hit|, For Layers 0, 6, 10, 14, 18, 22, 26, 29**

In Figure 9 is represented the Stepper (x, y) distribution of single Muon tracks at 3,4,5,10,20 and 50 GeV/c.

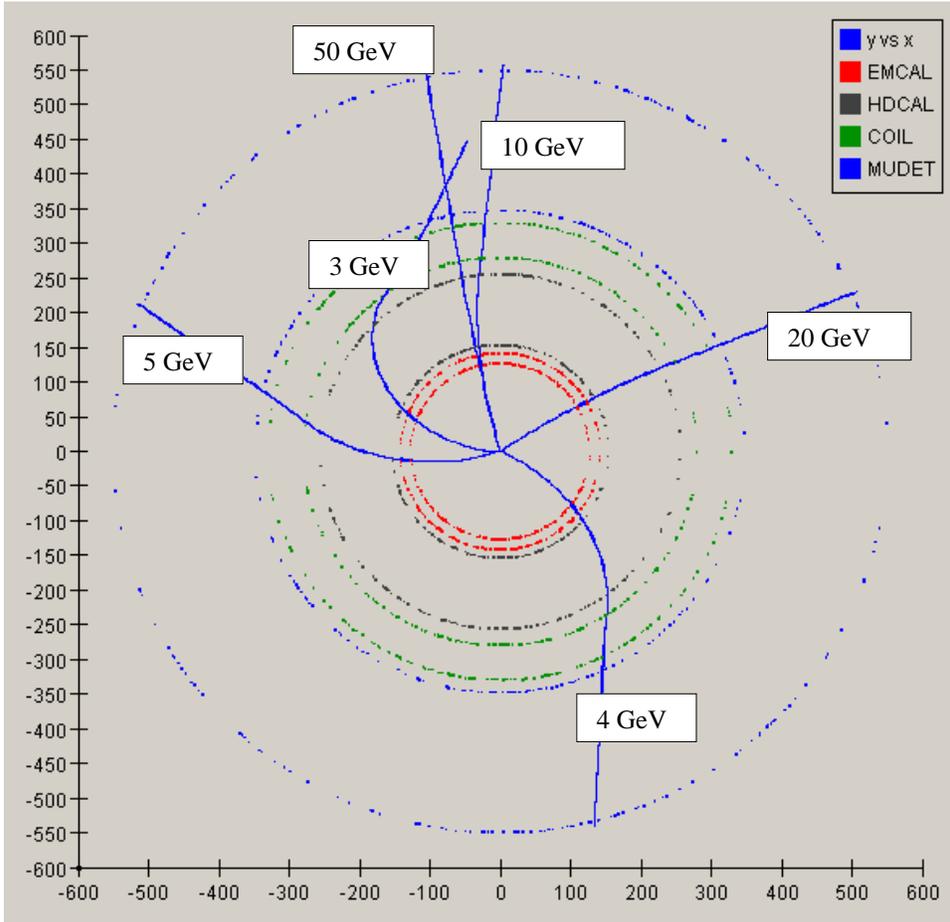

**Figure 10- The Stepper (x, y) Distribution For Single Muon Tracks at 3, 4, 5, 10, 20, 50 GeV**

The 3 GeV muon is curling in the magnetic field $B_z$ =5 Tesla in the tracker. The EMCAL boundaries are shown by two red circles; HDCAL is between the two black circles. Note the particle trajectory straightens in the coil, indicated by the two green circles, where the field drops and even changes sign. The trajectory is almost straight; in fact it is very slightly inverted in MUDET shown by the two blue circles. The 3 GeV/c particles do not get all the way through the MUDET, they stop somewhere around layer 16 in MUDET. The 4 GeV and 5GeV Muons have a very similar behavior but have enough energy to make it all the way through the MUDET.
 Above 5 GeV, the higher the particle momentum, the smaller the curling in the magnetic field as it appears above for the particles at 10, 20 and 50 GeV. The 50GeV muon looks even straight.

Following in Figures 10,11 are the (x, y) representations using the Single Event Display of the 3, 10 GeV Muons represented by the stepper in Figure 9.  One can see the trajectories produced by the stepper are a faithful representation of the events as seen by the event display.

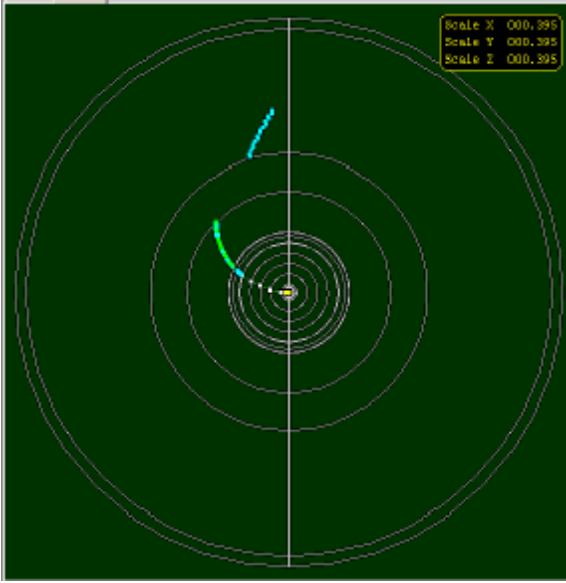

**Figure 11- The Event Display(x, y) Distribution of the 3 GeV Muon**

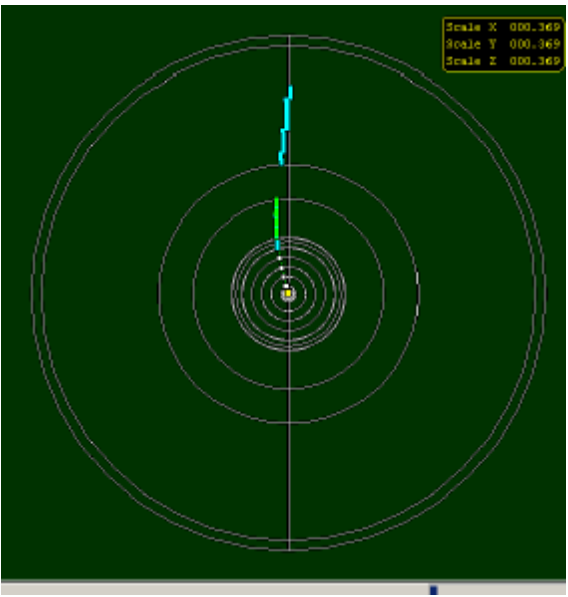

**Figure 12- The Event Display (x, y) Distribution of the 10 GeV Muon**

The evolution Through the Detector of the point ($p_x/p_{Int}$, $p_y/p_{Iint}$) ratio of the momentum components to it norm at Interaction Point ($p_{Int)}$ are shown for Single Muons of different energies using the stepper.
WARNING: The following plots are very different in behavior than the (x, y) track projections. They are complementary. One starts with the maximum momentum, e.g. 3 GeV/c, to which we normalize the 2 components of the momentum $p_x$ and $p_y$ in different elements of the Detector.

In the tracker $p_x$ and $p_y$ changes come only due to the magnetic field Bz but
$\sqrt{(p_x * p_x + p_y * p_y)} \sim p_{Int}$ stays constant, the material in the Tracker being negligible.
The point ($p_x/p_{Int}$, $p_y/p_{Int}$) is on the circle of radius 1, the norm of the momentum
Being almost unchanged. Then, in the calorimeters, the particle loses energy, and
therefore momentum in the material encountered. The new points ($p_x/p_{Int}$, $p_y/p_{Int}$) is
moving toward the center to a radius smaller than one. It loses even more energy in
HCAL than in ECAL (the $\Sigma((dE/dx)*\Delta x)$ being bigger) and goes on loosing energy in
the COIL and in the MUDET, but there, the magnetic field is inverted and smaller in
magnitude. Therefore the momentum starts high and ends up at or close to zero at 3
GeV/c, and the particles often stops in the MUDET, in about 20 layers or less .The point
($p_x/p_{Int}$, $p_y/p_{Int}$) ends up at the center of the circle.

The position (x, y), on the other hand, was starting at the Interaction Point at a radius
$r = \sqrt{(x*x + y*y)} \sim 0$ and increases to end up at a radius ~362 cm in the MUDET for a 3
GeV Muon. The 4 GeV/c muon is left with ~10% of its momentum at the border of the
instrumented Muon Detector, and the higher the Muon Momentum the smaller the
change in radius (the smaller the proportion of momentum loss), as can be seen in the
plot for 10GeV/c and 20 GeV/c Muons.

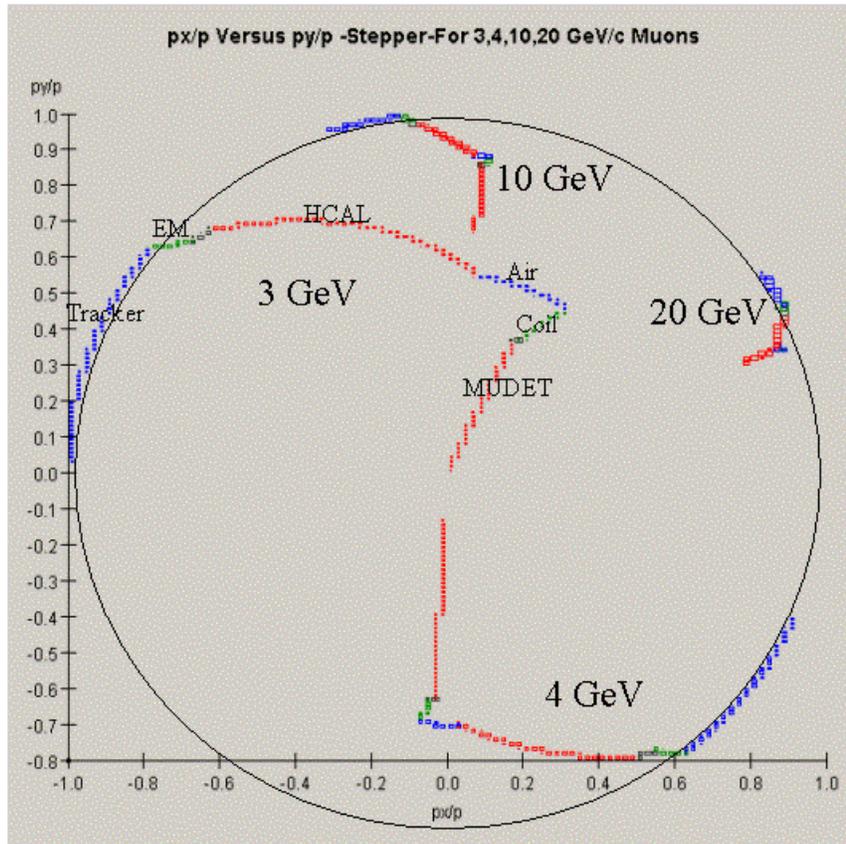

**Figure 13- Evolution Through The Detector Elements Of ($p_x/p_{Interaction}$, $p_y/p_{Interaction}$).**

**$p_{Interaction}$ Is The Norm Of The Momentum At Start At The Interaction Point**

The input values of the magnetic field fed into the stepper to represent the coil and
MUDET has been done to follow the simulated hits. The sudden change of slope of the particle at the entrance of the coil is quite un-realistic, we are expecting, a linear decrease, in a "real" world. We also expect a stronger bending back of the tracks in MUDET were the magnetic field is not correct. . This can be seen in the event display for single muons as well as in the track from the stepper shown 3 figures back, while representing x versus y for a Muon at 3, 4 5,10,20 and 50 GeV/c. This feature of the event generation needs to be corrected.

At higher energies the particle momentum does not reach zero at the end of the trajectory and the particle will go through the non- instrumented Iron in MUDET and even leave the detector.

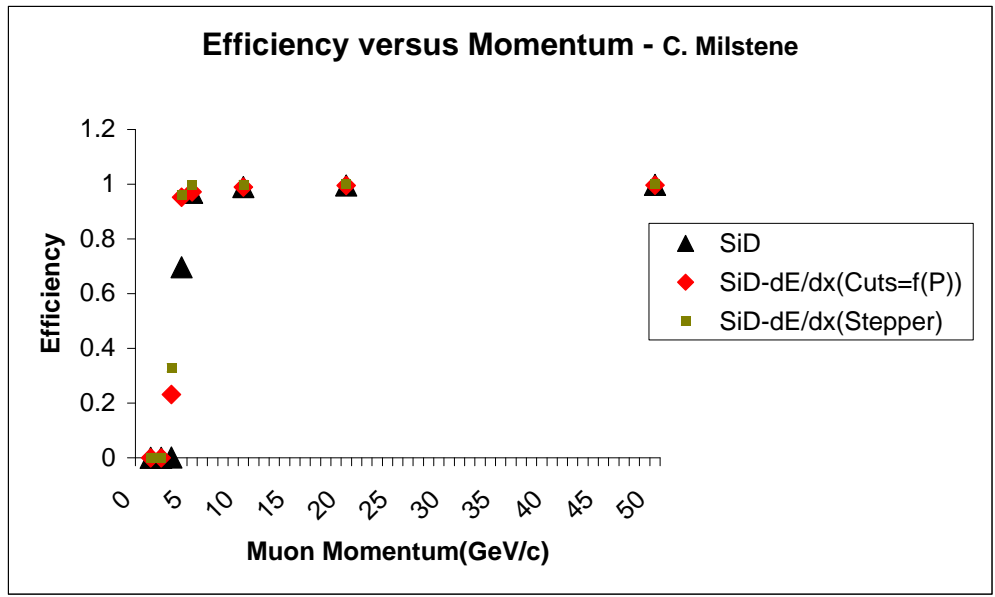

**Figure 14- Single Muon Efficiency As A Function of P (GeV/c)-Using The Swimmer Without/**

**With Account For An Ad-Hoc dE/dx And With The Stepper With The "Real" dE/dx .**

The Single Muons Efficiency is reported above as a function of p(GeV/c). The lower energy efficiency which improved in taking into account of the dE/dx by using a Momentum dependant improves now further with the stepper where the loss of energy due to dE/dx in subtracted step by step.
It is mostly at low Momenta that the improvement is more important
For 3 GeV/c Muons: The Efficiency went from:  0.6%   -> 23%        ->33%      Stepper
For 4 GeV/c Muons: The Efficiency went from:  ~70%    -> 95.2%   ->96.2%   Stepper
For 5 GeV/c Muons:  The Efficiency went from:  ~97%  -> ~97%    ->99.6%    Stepper
For 10GeV/c Muons: The Efficiency went from: 98.96% -> 98.96% ->99.98% Stepper
At higher energy the improvement is subtler.

# Conclusion

The stepper improves the low momentum muon identification efficiency further and one can show that it gets rid of part of the contamination, especially the low energy tracks which get stopped and are now flagged as such. Namely, one have seen that in jets, when the swimmer was used, the low momentum tracks which have a neighbor Muon are sometimes wrongly identified as Muons as well. They simply borrow hits in the MUDET from that neighbor Muon. When using the stepper those tracks are flagged as stopped And will not go through the same fate.
Many thanks to Philippe Gouffon for important comments and suggestions. Thanks to Fritz Dejongh for going through the process. The software of the stepper has benefited from advices and suggestions of Margherita Vittone Wiersma. Thanks are also due to NIU, Fermilab and A. Zaks for their continuous support.

# Bibliography
[1] hep\lcd\recon\muon\doc_files\MuonReco.pdf- A JAS Package for LCD Muon Reconstruction- by Richard Markeloff

# Appendix:

One assumes that changes in direction of the momentum between two steps is taken care entirely by the Magnetic Field. At step n the directions of P are

$$\frac{p_x(n)}{P(n)} = a1; \quad \frac{p_y(n)}{P(n)} = a2; \quad \frac{p_z(n)}{P(n)} = a3;$$

Due to the field Bz the particle changes direction to become at the end of the step
$$p_x' = p_x(n) + dp_x\_FromB; \quad p_y' = p_y(n) + dp_y\_FromB; \quad p_z' = p_z(n)$$
$$P' = \sqrt{p_x'^2 + p_y'^2 + p_z'^2}$$

and the new angles are given below and are the a'. The angles in the middle of the step are given by the a".

$$a_1' = \frac{p_x'}{P}; \quad a_2' = \frac{p_y'}{P}; \quad a_3' = \frac{p_z'}{P};$$
$$a_1'' = \frac{a_1 + a_1'}{2}; \quad a_2'' = \frac{a_2 + a_2'}{2}; \quad a_3'' = \frac{a_3 + a_3'}{2}$$

Using the angles in the middle of the step the changes in the component of the momentum due to material are given below.

<u>Remarks</u>:1) |P(n+1)|β|P(n)|, the particle loses energy in the material .

2) If the step is really small using the value of the angles in the middle of the step,( the a"), or at the beginning of the step,( the a), does not make a real difference in the calculation of the Δp due to dE/dx. In the 3 expressions above one replaces ΔP by the value above, where the a's replace the a", one gets,

$$\gamma_x(n) = \Delta p_x = a1 * \frac{E(n)}{P(n)} * Ct * \Delta s = \frac{p_x(n)}{P(n)} * \frac{E(n)}{P(n)} * Ct * \Delta s = \frac{p_x(n)}{P(n)} * \frac{E(n)}{P(n)} * (\frac{dE}{dx}) * \Delta s$$

$$\gamma_y(n) = \Delta p_y = a2 * \frac{E(n)}{P(n)} * Ct * \Delta s = \frac{p_y(n)}{P(n)} * \frac{E(n)}{P(n)} * Ct * \Delta s = \frac{p_y(n)}{P(n)} * \frac{E(n)}{P(n)} * (\frac{dE}{dx}) * \Delta s$$

$$\gamma_z(n) = \Delta pz = a3 * \frac{E(n)}{P(n)} * Ct * \Delta s = \frac{p_z(n)}{P(n)} * \frac{E(n)}{P(n)} * Ct * \Delta s = \frac{p_z(n)}{P(n)} * \frac{E(n)}{P(n)} * (\frac{dE}{dx}) * \Delta s$$

and therefore as given in the

$$\gamma_x(n) = \Delta p_x = E(n) * (\frac{dE}{dx}) * \frac{p_x(n)}{P^2(n)} * \Delta s = E(n) * (\frac{dE}{dx}) * \frac{p_x(n)}{p_x^2(n) + p_y^2(n) + p_z^2(n)} * \Delta s$$

$$\gamma_y(n) = \Delta p_y = E(n) * (\frac{dE}{dx}) * \frac{p_y(n)}{P^2(n)} * \Delta s = E(n) * (\frac{dE}{dx}) * \frac{p_y(n)}{p_x^2(n) + p_y^2(n) + p_z^2(n)} * \Delta s$$

$$\gamma_z(n) = \Delta p_z = E(n) * (\frac{dE}{dx}) * \frac{p_z(n)}{P^2(n)} * \Delta s = E(n) * (\frac{dE}{dx}) * \frac{p_z(n)}{p_x^2(n) + p_y^2(n) + p_z^2(n)} * \Delta s$$